\documentclass[10 pt,aps,prb,reprint,showpacs]{revtex4-1}

\usepackage[utf8x]{inputenc}	
\usepackage[T1]{fontenc}
\usepackage{color}
\usepackage{amsmath}
\usepackage{amsfonts}
\usepackage{latexsym}
\usepackage{amssymb}
\usepackage{bm}
 
\usepackage{graphics}
\usepackage{epsfig}
\usepackage{dcolumn}

\usepackage{hyperref}

\newcommand{\g}[1]{{\bf #1}}

\begin{document}

\title{Ferromagnetism in ${\text{U}}{\text{Ge}}_{2}$: A microscopic model}
 
\author{Marcin M. Wysoki\'nski}
\email{marcin.wysokinski@uj.edu.pl}
 
\author{Marcin Abram}
\email{marcin.abram@uj.edu.pl}

\author{Jozef Spa\l ek}
\email{ufspalek@if.uj.edu.pl}

\affiliation{Instytut Fizyki im. Mariana Smoluchowskiego$,$ Universytet Jagiello\'n{}ski$,$ 
Reymonta 4$,$ PL-30-059 Krak\'ow$,$ Poland}

\date{\today}

\begin{abstract}
 Anderson lattice model is used to rationalize the principal
 features of the heavy fermion compound UGe$_2$ by means of the 
 {\it generalized Gutzwiller approach} (the SGA method).
 This microscopic approach successfully reproduces magnetic and electronic
 properties of this material,
 in a qualitative agreement with experimental findings from the magnetization
measurements, the neutron scattering, and the de Haas--van Alphen oscillations.
Most importantly, it explains the appearance, sequence, 
character, and evolution in an applied magnetic field of the observed in UGe$_2$ 
ferro- and, para-magnetic phases as an effect of a competition between 
the $f$--$f$ electrons Coulomb interaction energy
and $f$--conduction electrons kinetic energy (hybridization).

\end{abstract}


\maketitle

{\it Introduction.} 
The discovery of the spin-triplet superconductivity (SC) 
inside the ferromagnetic (FM) phase of heavy-fermion compound UGe$_2$\ ~\cite{Saxena2000} 
sparked an intense discussion about the cause of such coexistence.
Although the spin-triplet paired phase has been known to 
appear in the condensed $^3$He\ ~\cite{Leggett1975} and most likely 
in Sr$_2$RuO$_4$\ ~\cite{Wysokinski2009,*WysokinskiK2012,*WysokinskiK2013}, until its
discovery in UGe$_2$ there was no convincing example for a strongly 
FM material hosting SC.

Specifically, the phase diagram for UGe$_2$ on the temperature--pressure (T--p)
plane contains both SC and two FM phases, with stronger and weaker 
magnetization \cite{Pfleiderer2002}, usually referred to as FM2 and FM1, respectively, 
as well as paramagnetic phase (PM), with the phase transitions between them of the 
1st order for low temperature, $T \lesssim 7K$\ ~\cite{Taufour2010}.
The FM--SC coexistence is strongly suggestive of a single mechanism 
based on magnetic correlations which is responsible for both FM and SC
appearance and thus should be treated on equal footing, as e.g. in UGe$_2$ both phases disappear 
at the same pressure \cite{Saxena2000, Pfleiderer2002}.
Another indication of the coupled nature of both phases is
that SC dome on the T--p plane coincides
with the phase transition between FM2 and FM1 \cite{Pfleiderer2002}.
Thus, we address here in detail the 
question of microscopic origin of the observed  
ferromagnetism, as it should bring us 
closer to determining the mechanism of
superconductivity. The question related to the 
inclusion of SC requires a separate 
 study \cite{Spalek2001,Zegrodnik2012} (see discussion at the end). 

 Experimental observations suggest that the ferromagnetism in UGe$_2$
has an itinerant nature \cite{Saxena2000,Pfleiderer2009,Aoki2012}
and is mediated by the uranium 5$f$ electrons \cite{Saxena2000,Huxley2001,Kernavanois2001}.
Delocalization of the 5$f$ electrons can be interpreted 
as resulting from hybridization of 5$f$ originally atomic 
states with those from conduction band \cite{Pfleiderer2009} derived from $p$ states due to
Ge and $d$-$s$ states due to U. 
This is supported by a~noticeable difference of 
the effective paramagnetic moment per uranium atom in this compound with respect 
to the corresponding atomic value for either $f^3$ or $f^2$ configurations \cite{Saxena2000},
as well as from a fractional value of the magnetization relative to the moment saturation.
This means that the Hund's rule coupling in the atomic sense is broken, and 
the itineracy of 5$f$ electrons is the source of the band ferromagnetism 
in which Hund's ferromagnetic interaction plays a role in combination
with the stronger intra-atomic Coulomb interaction.
This also means that the $f$-electron orbital degeneracy is not essential, but the role 
of the hybridization is.
    
Apart from other theories concerning origin of FM in considered class of materials 
\cite{Hirohashi2004,Kirkpatrick2005} there exists \cite{Sandeman2003} 
a phenomenological rationalization of the magnetic
properties within a rigid-band Stoner approach, which requires introduction of an 
ad hoc two-peaked structure of density of states (DOS) near the Fermi surface (FS). Our purpose is to 
invoke a microscopic description starting from the Anderson-lattice model (ALM) which 
 is appropriately adapted to the heavy-fermion compound UGe$_2$. 
This comprises a relatively simple quasi-two-dimensional 
 electronic structure \cite{Shick2001, Tran2004, Onuki1992}. 
 From such starting point an effective non-rigid two-band 
 description arises naturally and allows for a detailed rationalization
 of magnetic and electronic properties, at least on a semiquantitative level. 
 Additionally, as the correlations among 5$f$ electrons
 are sizable, an emergence of the Stoner-like picture 
 of FM can be accounted for only
 with inclusion specific features coming from the electronic correlations. Although the resulting
 rationalization of the physical properties is semiquantitative 
 in nature, it provides in our view a coherent  picture of a number of properties
 \cite{Pfleiderer2002, Settai2002, Terashima2001,Huxley2001, Kernavanois2001}.   

{\it Model.}
We solve ALM by means of a~variational treatment with the Gutzwiller 
wave function, ${|\psi_G\rangle=\prod_{\g i}\hat P_{\g i}|\psi_0\rangle}$, 
where $\hat P_{\g i}$ is the operator projecting out part of  
double occupancies from the uncorrelated groundstate $|\psi_0\rangle$ at site $\g i$.
We have extended the standard approach \cite{Vollhardt1984,Rice1985,Fazekas1987} 
to the {\it statistically consistent form}
 \cite{ *[{Approach derived in }] [{ was used successfully in many cases, i.e. }] 
 SGA,*Jedrak2010,*Jedrak2011,*Kaczmarczyk2011,*WysokinskiVietri,*Howczak2012,*Abram2013,
 *Howczak2013,*Wysokinski2014} (SGA method). 
 Explicitly, we start with the ALM Hamiltonian, with an applied magnetic field 
introduced via the Zeeman term ($h \equiv \frac{1}{2} g \mu_B H$), i.e.,
\begin{equation}
 \begin{split}
 \mathcal{\hat H}-\mu\hat N&={\sum_{\g i,\g j,\sigma}}'  t_{\g i \g j}\hat c_{\g i,\sigma}^\dagger\hat c_{\g j,\sigma}
-\sum_{\g i,\sigma}(\mu+\sigma h)\hat n^c_{\g i,\sigma}\\
&+ \sum_{\g i,\sigma}(\epsilon_f-\mu-\sigma h)\hat n^f_{\g i,\sigma}+ 
U\sum_{\g i} \hat n^f_{\g i,\uparrow} \hat n^f_{\g i,\downarrow}\\
&+ V\sum_{\g i,\sigma}(\hat f_{\g i,\sigma}^\dagger
\hat c_{\g i,\sigma}+\hat c_{\g i,\sigma}^\dagger\hat f_{\g i,\sigma}),\label{Ho}
\end{split}
\end{equation}
where primed sum denotes summation over all lattice sites 
$\g i \neq \g j$, $\hat f$ and $\hat c$ are operators related to 
$f$- and $c$- orbitals respectively, with spin $\sigma=\uparrow, \downarrow$. 
We have also defined the total number of electrons operator as $\hat N$, 
and for the respective orbitals and spins as 
$\hat n^f_{\g i,\sigma} \equiv \hat f_{\g i,\sigma}^\dagger \hat f_{\g i,\sigma}$,
$\hat n^c_{\g i,\sigma} \equiv \hat c_{\g i,\sigma}^\dagger \hat c_{\g i,\sigma}$.
In our model, we consider finite intra-$f$-orbital Coulomb 
interaction $U$, the on-site inter-orbital hybridization
$V<0$, the hopping amplitude between the first ($t$), 
and the second ($t'=0.25\ |t|$), nearest neighboring sites, and the 
atomic level for $f$-states placed at $\epsilon_f=-3|t|$.
In the following $|t|$ is used as energy unit.

First, we would like to evaluate the ground-state energy,
${E_G\equiv\langle \psi_G\mid\mathcal{\hat H}\mid\psi_G \rangle / \langle\psi_G\mid\psi_G\rangle}$.\!
Applying the usual proce\-du\-re \cite{Rice1985,Fazekas1987}, called Gutzwiller approximation (GA),
we simplify the projection to the local sites on which the operators from (\ref{Ho}) act. 
In that manner one obtains the effective single-particle Hamiltonian in a momentum space with 
renormalized hybridization by the 
Gutzwiller narrowing factor $q_\sigma$~\cite{Vollhardt1984}, namely
\begin{equation}
\begin{gathered}
\mathcal{\hat H}_{GA}\equiv\sum_{\g{k},\sigma}  \Psi^\dagger
\begin{pmatrix}
 \epsilon_{\g{k}}^{c}-\sigma h-\mu&\sqrt{q_\sigma}\,V\\
\sqrt{q_\sigma}\,V& \epsilon_{f}-\sigma h-\mu   \\
\end{pmatrix}
\Psi +\Lambda Ud^2,
\end{gathered}\label{HGA}
\end{equation}
where we have defined $\Psi^\dagger\equiv(\hat c_{\g{k},\sigma}^\dagger, \hat f_{\g{k},\sigma}^\dagger)$, 
$\Lambda$ denotes number of lattice sites, and $d^2$ is the probability of 
having doubly occupied $f$-orbital that we optimize variationally. 
In order to ensure that variationally calculated polarization 
and the $f$-level occupancy would coincide 
with those coming from the self-consistent procedure \cite{SGA}, we modify 
our effective Hamiltonian (\ref{HGA}) by introducing additional 
constraints on the polarization ($m_f$) and the number ($n_f$) of 
$f$-electron states via the Lagrange-multiplier method. The effective Hamiltonian
with the constraints takes now the form
\begin{multline}
 \mathcal{\hat H}_{SGA} \equiv  \\
 \mathcal{\hat H}_{GA} - \lambda^f_n \Big( \sum_{\g{k},\sigma}\hat n^f_{\g{k},\sigma}- \Lambda n_f \Big)
 -\lambda^f_m \Big( \sum_{\g{k},\sigma}\sigma\hat n^f_{\g{k},\sigma} -  \Lambda m_f \Big) \vspace{3pt}\\
= \sum_{\g{k},\sigma}  \Psi^\dagger
\begin{pmatrix}
 \epsilon_{\g{k}}^{c}-\sigma h-\mu&\sqrt{q_\sigma}V \vspace{3pt}\\
 \sqrt{q_\sigma}V& \epsilon_{f}-\sigma (h+\lambda_{m}^f)-\lambda_{n}^f-\mu   \\ 
\end{pmatrix}
\Psi \\
+\Lambda( Ud^2 +\lambda_{n}^fn_f +\lambda_{m}^fm_f).\vspace{3pt}
\label{HSGA}
\end{multline}
Those constraint parameters $\lambda_n^f$ and $\lambda_m^f$ are also determined variationally.
They play a role of nonlinear self-consistent fields acting on the charge and the spin degrees
of freedom respectively. Diagonalization of (\ref{HSGA}) in this spatially homogeneous case leads to 
four branches of eigenenergies, $E_{{\bf k}\sigma}^\pm$ representing two spin-split hybridized bands $E^{\pm}$.
In order to determine the equilibrium properties of the system, we need
to find the minimum of the generalized Landau grand-potential functional $\mathcal{F}$, 
\begin{equation}
\begin{split}
\frac{ \mathcal{F}}{\Lambda}  = & -\frac{1}{\Lambda\beta} \sum_{{\bf k} \sigma b} \ln[1+e^{-
\beta E_{{\bf k}\sigma}^b}]\\
&+ (\lambda_{n}^fn_f+\lambda_{m}^fm_f+Ud^2),
\label{5}
\end{split}
\end{equation}
where $b=\pm$.
Effectively, it leads to the set of five nonlinear equations, 
$\frac{\partial \mathcal{F}}{\partial\vec \lambda}=0$ 
for $\vec \lambda\equiv\{d,n_f,m_f,\lambda_n^f,\lambda_m^f\}$.
However, due to the fact that the total number of electrons 
remains constant when the pressure 
or magnetic field is applied, we need to satisfy equation 
for the chemical potential $\mu$ via the condition 
\begin{equation}
 n=\frac{1}{\Lambda}\sum_{{\bf k}b\sigma}  f ( E_{{\bf k} \sigma}^{b} ),
\end{equation}
with $f$ being the Fermi distribution.  
The equilibrium thermodynamic potential functional defines also the ground state energy, 
${E_G=\mathcal{F}|_0+\Lambda\mu_0 n}$, where subscript $'0'$ denotes the optimal values.
After carrying out the minimization, we can also calculate total spin polarization from
\begin{equation}
 m\equiv m_c+m_f=\frac{1}{\Lambda}\sum_{{\bf k}b\sigma}  \sigma f ( E_{{\bf k} \sigma}^{b}).
\end{equation}

The numerical calculations with the precision of at least $10^{-7}$
were carried out for a~two dimensional, square lattice of $\Lambda = 512 \times 512$ size,
and for low temperatures ${\beta\equiv 1/{k_B T} \geqslant 1500}$,
emulating the ${T\rightarrow0}$ limit.

\begin{center}\vspace{-0.5cm}
  \begin{figure}
   \includegraphics[width=0.5\textwidth]{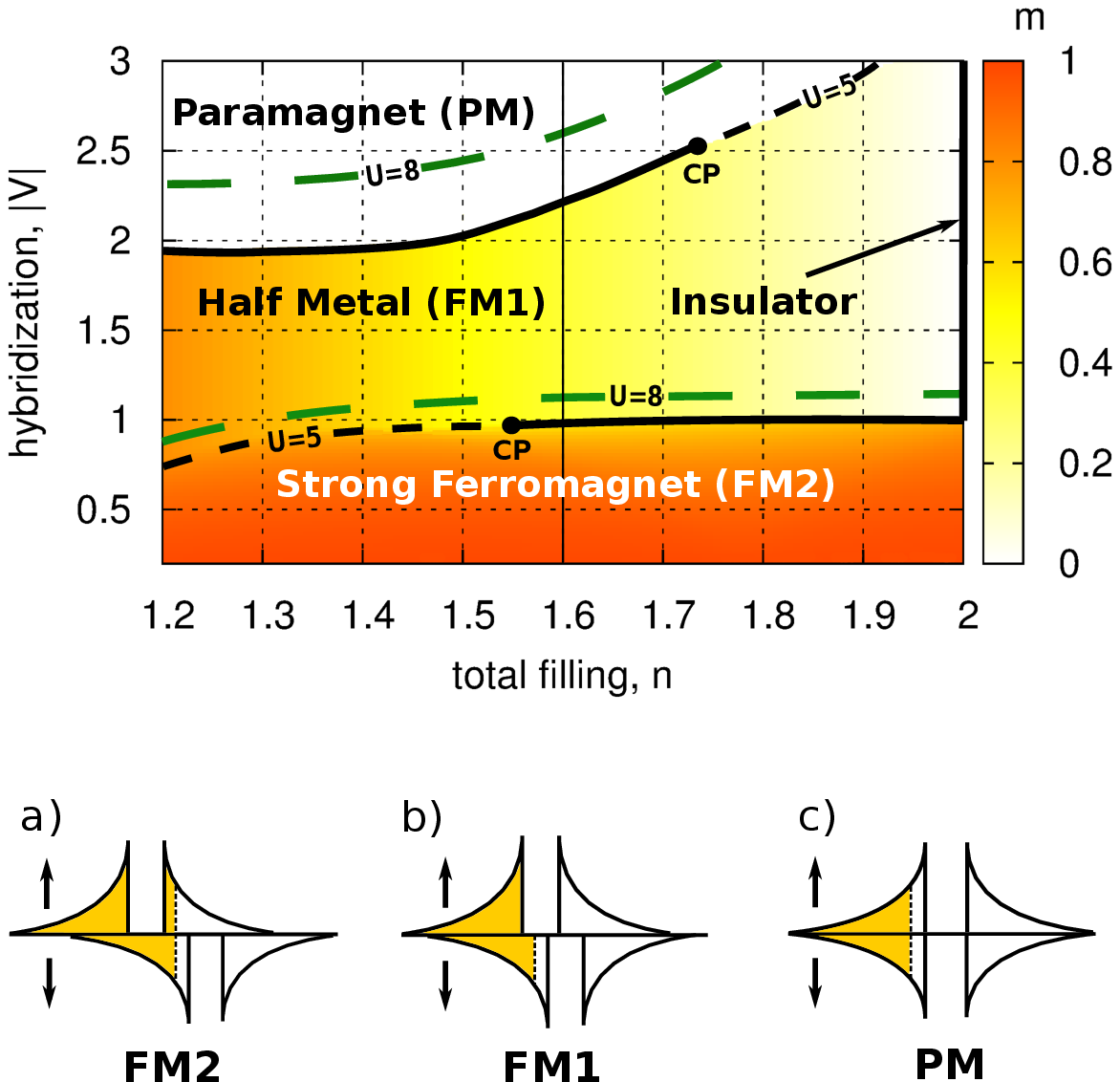}
   \caption{(Color online): (top) Phase diagram on plane 
   total filling--hybridization strength for the zero field, 
   containing both FM and PM phases for $U=5$. 
   Color scale denotes total spin polarization, $m$. 
   Phases are divided by the dashed and the solid lines. Dashed lines denote the 2nd order transition, whereas
   the solid -- the 1st order transition with the critical points, CP. Fine-dashed 
   lines mark how the phase borders would change for $U=8$.
   (a--c) Figures a--c depict a schematic
   spin-resolved density of states corresponding to the phase sequence appearing along
   the solid vertical line (from bottom to top).}\label{nv}
  \end{figure}
 \end{center}\vspace{-0.3cm}
 
\begin{center}
  \begin{figure}
\begin{minipage}{0.3\textwidth}
    \includegraphics[width=\textwidth]{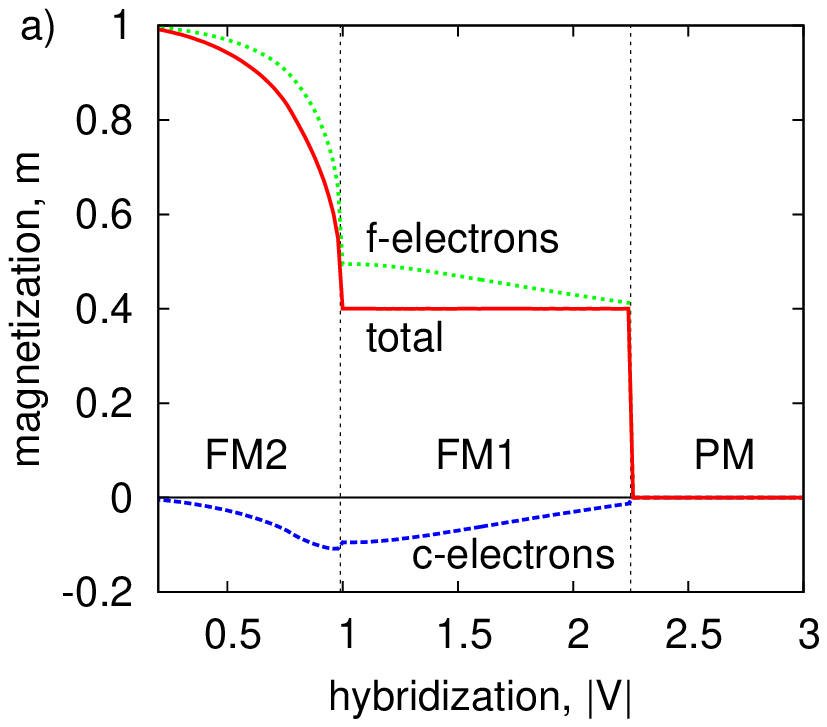}
\end{minipage}\hspace{-12pt}
\begin{minipage}{0.175\textwidth}
   \includegraphics[width=0.85\textwidth]{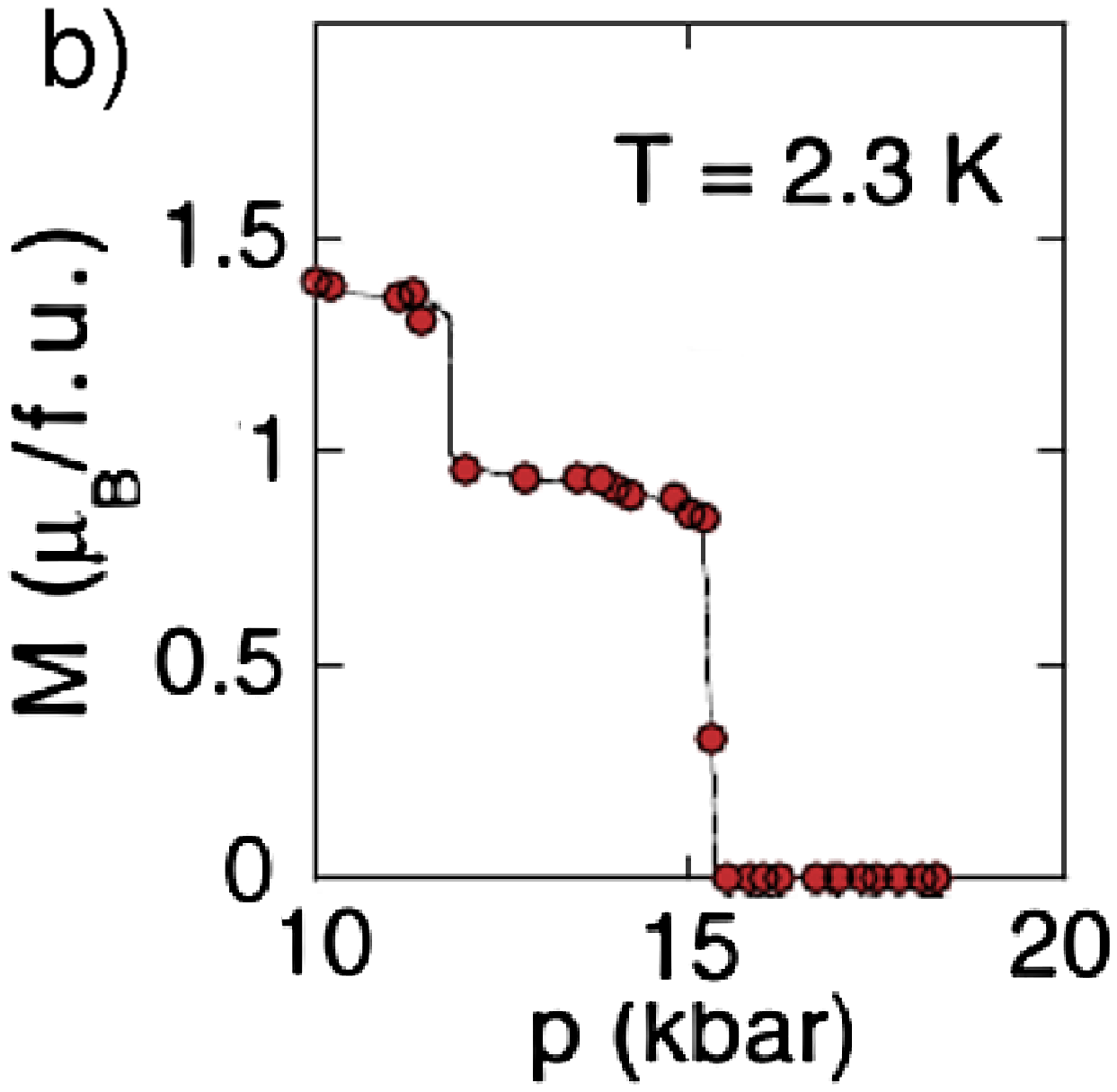}
   \includegraphics[width=1.0\textwidth]{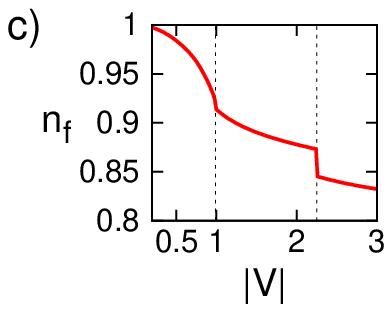}
\end{minipage}
   \includegraphics[width=0.45\textwidth]{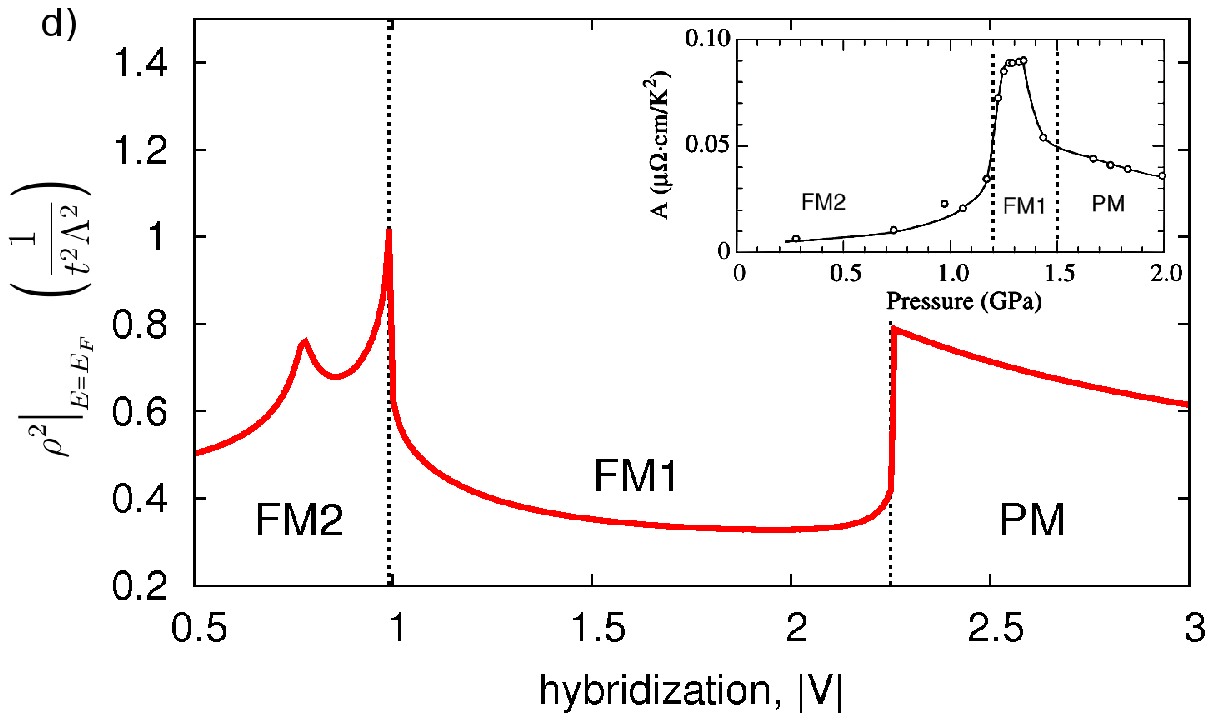}
   \caption{(Color online): (a) Magnetization as a~function of hybridization strength for the
   band filling $n=1.6$, and the Coulomb repulsion $U=5$.
   Both phase transitions induced by the hybridization change are of the 1st order. 
   (b) Corresponding experimental results from Ref. \onlinecite{Pfleiderer2002}.
   (c) $f$-orbital filling as a~function of hybridization.
   (d) Square of DOS at the Fermi level versus $|V|$ through the phase sequence. 
   Inset: Experimentally  measured $T^2$-term coefficient $A$ of the resistivity versus
   pressure from Ref. \onlinecite{Settai2002}.
    }\label{mag}
  \end{figure}
 \end{center}

{\it Results.} 
First, we analyze FM and PM 
solutions in the field absence. In Figure~\ref{nv}
we draw phase diagram on total filling--hybridization strength plane. 
For low hybridization, FM phases are favored due to the 
negative balance between increase of the kinetic and decrease
of the Coulomb energies, caused by a relative shift of the spin-resolved DOS.
This is visualized by the diminution of the spin-subband
overlap up to the FS -- cf.\ Figs.~\ref{nv}{\it a} and~\ref{nv}{\it b}.
The appearance of a 
spontaneous polarization, as a result of a competition between the kinetic and 
the Coulomb energies, is in fact the feature of the Stoner mechanism for the band FM onset. 
In comparison to the usual single-band (e.g. Hubbard) model, we
can distinguish in a~natural manner between the two FM 
phases. The first (FM1) appears when the chemical 
potential is placed in the hybridization gap, between the 
spin subbands of the lower hybridized band which is characterized also by the  
magnetization equal to $m=2-n$ (cf.\ Fig.~\ref{nv}{\it b}). 
In that situation, only the spin-minority carriers are present at and near FS.
The second phase (FM2) emerges when we further lower the hybridization 
and thus the chemical potential enters the
 majority spin-subband DOS (cf.\ Fig.~\ref{nv}{\it a}), giving 
rise to a~step (discontinuous) increase in magnetization (cf.\ Fig.~\ref{mag}{\it a}).
In the limit of strong hybridization, for a~fixed total filling,
when the correlations weaken due to lowering of $f$-orbital 
average occupancy ($n_f\lesssim0.85$, c.f. Fig.~\ref{mag}{\it c}),
the kinetic energy gain outbalances a~subsequent reduction of the average
Coulomb interaction and PM phase is energetically favorable.
Similar mechanism for the formation of FM and in 
particular, characterization of phases, was studied before in 
Refs. \onlinecite{Doradzinski1997,Doradzinski1998,Howczak2012,Kubo2013}.

We presume that the main effect of the pressure exerted on the 
material can be modeled by a concomitant strengthening 
of the hybridization amplitude. Thus, from Fig.~\ref{nv} it can be 
seen that for the total filling $n$ in the range
$1.55$--$1.75$, the sequence of phases
and the order of the transitions are the same, as those found experimentally 
for UGe$_2$ by increasing the pressure \cite{Pfleiderer2002,Pfleiderer2009}. 
As a representative band filling we have selected $n=1.6$ marked by the vertical line in Fig.~\ref{nv}.
In fact, as we compare the magnetization versus 
hybridization along the traced line (cf. Fig. \ref{mag}{\it a}) with
the corresponding experimental data \cite{Pfleiderer2002} (cf. Fig. \ref{mag}{\it b}) we find 
good qualitative resemblance. Moreover, the magnetization differentiation among
the orbitals (cf. Fig. \ref{mag}{\it a}) is in an agreement with the neutron scattering 
data \cite{Kernavanois2001, Huxley2001} at ambient pressure (in our model $|V|\simeq 0.5$), 
where it was found that almost exclusively electrons from uranium atoms ($f$-orbital) 
contribute to the ferromagnetism.
In our Stoner-like picture it is resulting from the fact that the competition
between Coulomb repulsion and hybridization-induced itineracy concerns
mainly the $f$ electrons.
Furthermore, as for low hybridization (in FM2 phase) we obtain a~small 
compensating polarization due to the $c$-electrons, we suggest that the experimentally 
observed small negative magnetization between the uranium atoms at ambient pressure \cite{Kernavanois2001}
may come from the delocalized cloud of conduction electrons.
 \vspace{-0.9cm}
 
Our microscopic description of the phase transitions induced by
the change of the FS topology compares also favorably with the electronic-state features 
of UGe$_2$ derived from de Haas van Alphen oscillations \cite{Settai2002,Terashima2001}. 
In Ref. \onlinecite{Settai2002} it is suggested, 
that the majority spin FS disappears in the FM1 phase, in complete accord with the
character of DOS presented in the Fig.~\ref{nv}{\it b}. 
We also reproduce the feature of an abrupt change of the 
FS at the FM1--PM phase transition 
\cite{Terashima2001,Settai2002} (cf. Fig. \ref{mag}{\it d}). 
Namely, it corresponds here to the step change of the chemical 
potential position merging into both bands.
Furthermore, in the experimental data at the metamagnetic 
phase transition there is observed significant enhancement
of the quasiparticle mass renormalization \cite{Terashima2001}.
As it is proportional to the DOS at Fermi level, in
Fig. \ref{mag}{\it d} we provide the corresponding behavior, 
which can be understood within our model by the chemical potential 
crossing high hybridization peak in the majority spin subband.
The transition leads then to a step change of FS 
only in the majority spin subband, while the minority subband 
evolves rather continuously, what is also seen experimentally \cite{Terashima2001}.
 \begin{center}\vspace{-0.5cm}
  \begin{figure}
   \includegraphics[width=0.5\textwidth]{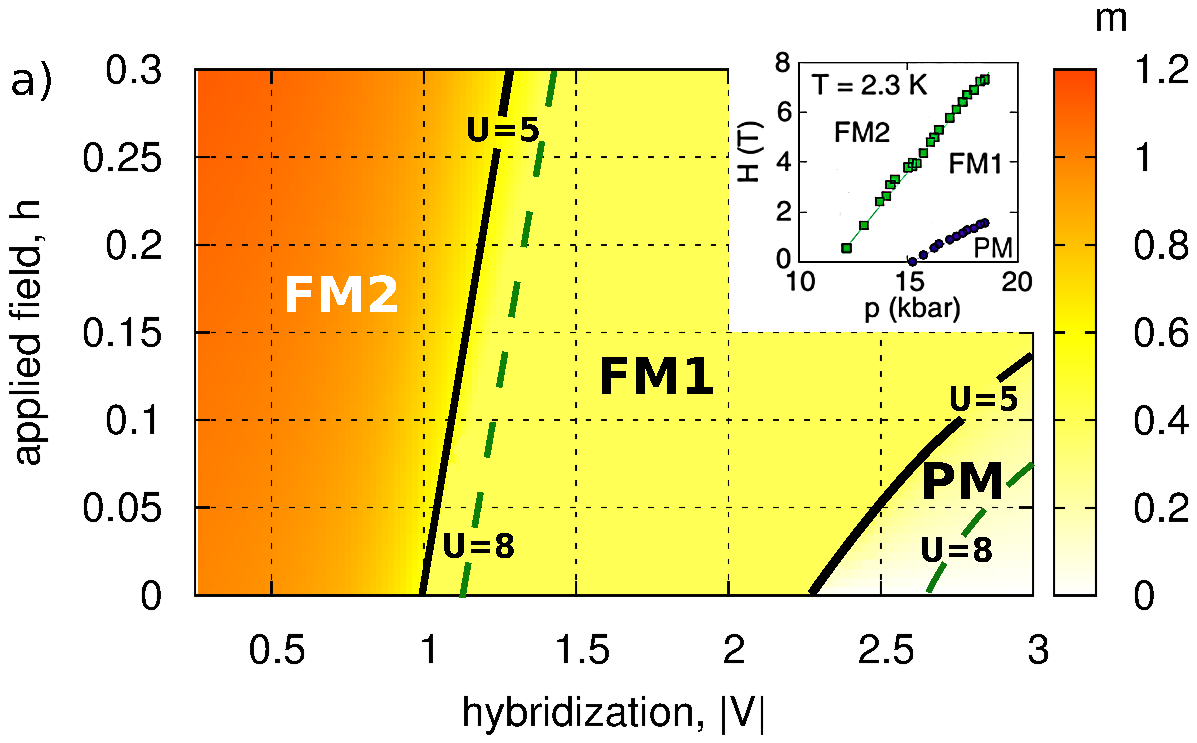}\\
   \includegraphics[width=0.235\textwidth]{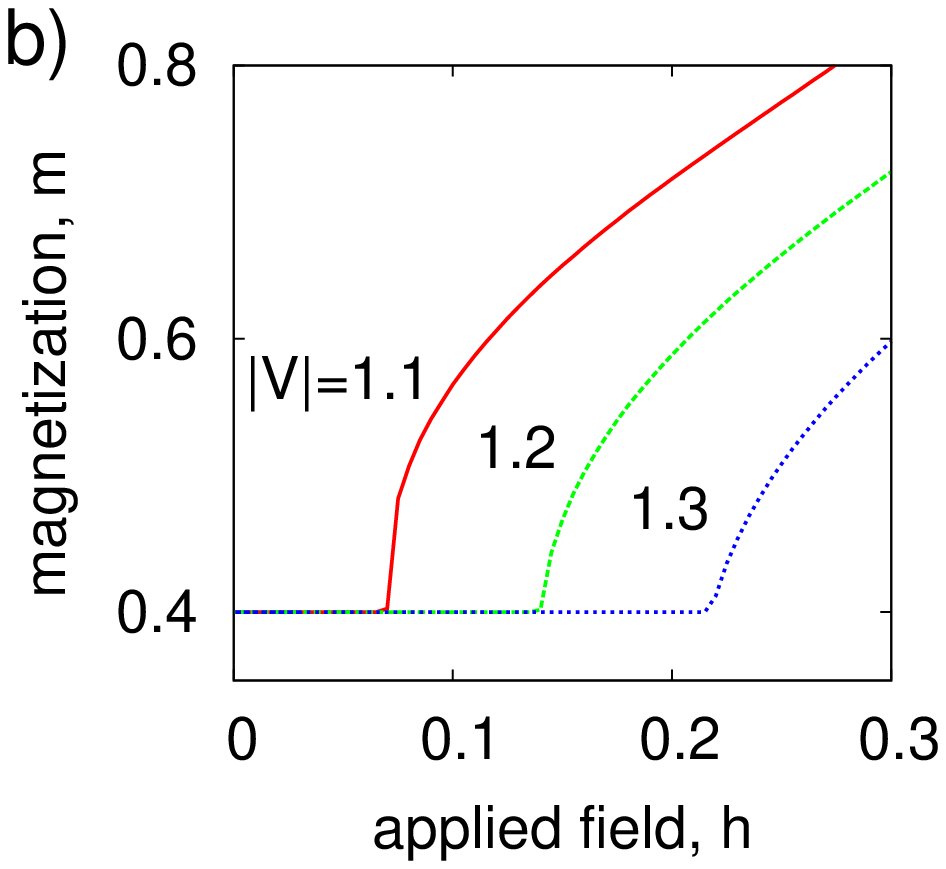}
   \includegraphics[width=0.235\textwidth]{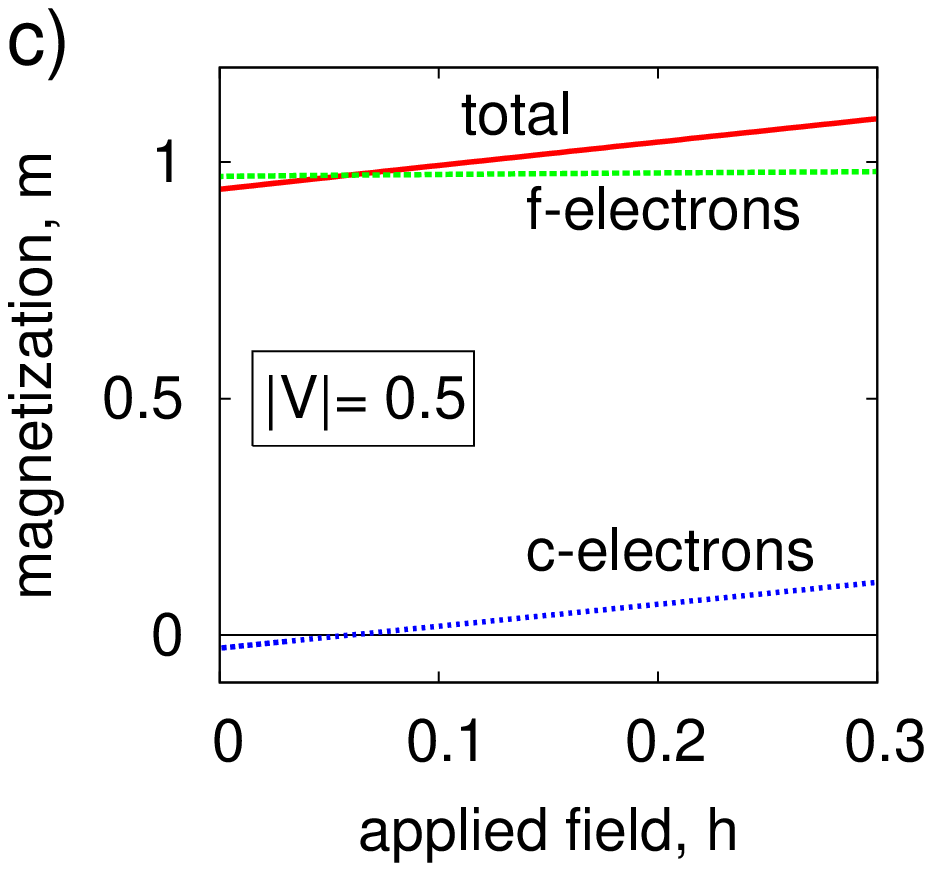}
   \caption{(Color online): (a) Phase diagram on the applied field--hybridization strength plane for $n=1.6$
   and $U=5$. Color scale denotes total spin polarization. The dashed 
   lines mark the phase stability thresholds for $U=8$. 
   In the inset we show experimental results \cite{Pfleiderer2002}. 
   (b) Magnetization versus applied field for selected hybridization strengths 
   when system is entering into FM1 to FM2 phase-transition regime.
   (c) Evolution of orbital-resolved magnetization with the 
   field for low hybridization, $|V|=0.5$ (mimicking ambient pressure). 
   Note the very small $c$-electron polarization up to $h\simeq0.1$.
   }\label{vh}\label{pole}
  \end{figure}
 \end{center}\vspace{-0.3cm}

For the sake of completeness, we have shown in the inset in 
Fig. \ref{mag}{\it d} the
pressure dependence of the $T^2$ term of resistivity \cite{Settai2002} 
as it should have roughly the same dependence as squared
DOS at the FS, versus $|V|$ (we assume that the Kadowaki-Woods scaling holds).
However, the jump that we obtain at the FM1 - PM transition has not been
observed in the resistivity measurements \cite{Settai2002}.

In the applied field, our model is also in good agreement with
 available experimental data for UGe$_2$. In Figure~\ref{vh}{\it a} we display phase 
 diagram on hybridization--applied-magnetic-field plane that 
 corresponds to that determined experimentally \cite{Pfleiderer2002} (cf. Fig.~\ref{vh}{\it a} inset).
 Similarly as in Ref. \onlinecite{Pfleiderer2002}, the magnetization at 
 the phase transition between FM1 and FM2 
 triggered by the applied magnetic field, starts from the same baseline, 
 independently of the
 hybridization strength (cf.\ Fig.~\ref{pole}{\it b}).
 However, one should note that, due to the fact that pressure changes not only 
 the hybridization magnitude, but also other microscopic parameters, we are not able 
to reproduce the magnetization cascade with the increasing magnetic 
field when crossing the transitions.
 
The next feature found in UGe$_2$ at ambient pressure is an initial lack 
of measurable polarization on the germanium atoms with the increasing magnetic field,
as inferred from the neutron scattering data \cite{Huxley2001}.
In our model we find a similar trend.
For low hybridization ($|V|\simeq0.5$ emulating
ambient pressure), $c$-electrons polarization increases slowly, and
even up to $h\approx0.1$ it is negligible (cf.\ Fig.~\ref{vh}{\it c}).

{\it Remarks.}
With the simple but powerful technique based on the generalized Gutzwiller ansatz (SGA method), 
applied to the Anderson lattice model, we have constructed a
microscopic model of FM in UGe$_2$.
Namely, we are able to reproduce main experimental features observed at low temperature,
by applying either pressure or magnetic field (cf. Figs. \ref{mag}{\it a} and \ref{vh}{\it a}).
FM properties can be rationalized within the simplest 
hybridized two-orbital model, without taking into account
the $f$-orbital degeneracy, i.e., by effectively incorporating both the Coulomb 
and the Hund's-rule interaction
into an effective interaction $U$, as would be also the case in the Hartree-Fock approximation \cite{Zegrodnik2012}.

To determine the stability of SC inside 
FM phase, the present approach should be extended to account for the 
Hund's-rule interaction explicitly what can be crucial for a formation of the 
unconventional triplet SC \cite{Spalek2001,Zegrodnik2012,Zegrodnik2013,Zegrodnik2014}. 
If this is the case, it can be triggered even by a purely 
repulsive Coulomb interaction in conjunction with
the residual Hund's rule coupling, as discussed in Refs. \onlinecite{Zegrodnik2013,Zegrodnik2014}.
This issue requires a separate analysis.
Another path for discussing the coexistence of SC with FM could be 
going beyond the Gutzwiller approximation,
 where we account also for the more distant correlations when determining
the effective Hamiltonian \cite{Kaczmarczyk2013, Kaczmarczyk2014}.
Here, the central question is whether the spin triplet pairing should be treated 
on the same footing as ferromagnetism, i.e., appears already in direct space formulation
\cite{Spalek2001,Zegrodnik2012,Zegrodnik2013,Zegrodnik2014} or is it mediated by collective 
spin fluctuations in ferromagnetic phase \cite{Fay1980,Kirkpatrick2001,Millis2001,Sandeman2003} 
among already well defined quasiparticles. 
A crossover from the latter to the former approach
is expected to take place with the increasing strength of the repulsive Coulomb interaction U.

{\it Acknowledgements.} The work has been partially supported by the Foundation for 
Polish Science (FNP) under the Grant TEAM, as well as by the 
National Science Centre (NCN) under the Grant MAESTRO, No. DEC-2012/04/A/ST3/00342.
We would like to thank J. Kaczmarczyk for discussions and critical reading of the manuscript.

\end{document}